\documentclass[prl,superscriptaddress, preprint]{revtex4}
\usepackage[utf8x]{inputenc}
\usepackage{graphicx}
\usepackage{textcomp}
\usepackage{bbm}
\usepackage{listings}
\usepackage{amssymb}
\usepackage{amsmath}
\usepackage{times}
\usepackage{txfonts}
\begin{document}

\title{Cortical columns for quick brains}
\author{Ralph L. Stoop$^1$, Victor Saase$^2$, Clemens Wagner$^{3}$, Britta Stoop$^2$, Ruedi Stoop$^2$}, 
\address{Physics Department, University of Basel, Klingelbergstrasse 82, CH-4056 Basel \\
$^2$Institute of Neuroinformatics, University of Zurich and ETH Zurich\\
Winterthurerstr. 190, 8057 Zurich \\
$^3$Fachdidaktik Physik, ETH Zurich, Schafmattstrasse 16,
CH-8093 Zurich\\
Email: ruedi@ini.phys.ethz.ch}

\begin{abstract}
It is widely believed that the particular wiring observed within cortical columns boosts neural computation. We use rewiring of neural networks performing real-world cognitive tasks to study the validity of this argument. In a vast survey of wirings within the column we detect, however, no traces of the proposed effect. It is on the mesoscopic inter-columnar scale that the existence of columns - largely irrespective of their inner organization - enhances the speed of information transfer and minimizes the total wiring length required to bind the distributed columnar computations towards spatio-temporally coherent results.
%Rather, the columnar architecture's purpose may be to minimize the network's total wiring length. %Rather, they corroborate earlier findings by Jaeger that Liquid State Machine performance depends little on the degree of recurrent connectivity.
\end{abstract}

\maketitle

\section*{Author summary}
{\it Cortical columns with their wiring substructures are a biological fact known for more than one hundred years. Yet, their function and computational role have largely remained unexplained. Recent computational work has put forward that this organization provides a direct computational advantage. A corresponding conclusion was drawn from an artificial computational context using detailed simulations of neurons and their interactions, but up to date, this finding has not been validated in the context of real-world applications. While covering an extremely wide range of network configurations, neuron and network models and cognitive tasks, we find no sign of a computational boost triggered by the inner-columnar template. 
We provide an alternative explanation: Doubly fractal connectivity laws with exponents as found in the cortex will optimize the speed of signal transfer in the cortex using minimal wiring length. While such laws do not necessarily imply cortical columns, a self-similarity of columnar structures would be a simple way to implement this connectivity. %Our analysis is based on neural networks of biologically detailed node elements and on coupled map lattice models, for the microscopic (inner-columnar) and the mesoscopic (inter-columnar) description, respectively. 
}

\section*{INTRODUCTION}
Towards the turn of the 19'th century, J. P. M\"uller, E. du Bois-Reymond and H. von Helmholtz \cite{Mueller} discovered that neurons are electrically excitable and this predictably affects the electrical state of connected neurons. Shortly after, Golgi and Ram\^on y Cajal \cite{GolgiCajal} provided their Nobel-prize winning description of neuronal and cortical architecture, revealing in the case of the human neocortex striking columnar structures divided into six layers. Ever since this discovery it has remained a question to what extent neuronal physics and the cortical architecture could account for the exquisite properties of the human brain, at least within the scope limited by G\"odel's theorem \cite{Goedel}. Recent attempts at solving this problem concentrate on physically building the brain, using electrical neurons organized according to the cortical blueprint (e.g., \cite{bluebrain}). 
Motivated by typical construction constraints encountered in chip making, we explore three potential benefits of the cortical wiring template. On the inner-columnar scale we measure the effect that the columnar wiring has on real-world pattern recognition tasks in a framework \cite{Jaeger07,Lukosevicius09} that permits to measure recognition rates without compromising the columnar wiring by the learning process. On the inter-columnar scale, the effects on the speed of information transport and computation are analyzed in a framework \cite{Kaneko, Bunimovich} that offers analytical methods with results valid for sufficiently general situations. Our approach is to start from networks in which details of cortical architecture are implemented. Using various rewiring processes, we move away from the biological blueprint,  measuring the effects of the removal. On the columnar level, our work parallels some investigations performed in Ref.~\cite{Haeusler07} in a more abstract setting, where a beneficial influence of microcircuit structures on computation was reported.

\section*{BIOLOGICAL DATA, COLUMN MODELS, MODELS OF NETWORKS OF COLUMNS}

{\bf Biological data:} We use the biological data collected by Roerig et al. \cite{Roerig02} for a similar context. A Log-Log-plot adaption of their original data (see Fig.~1) evidences a break between two decay laws at the scale indicated in Fig.~1 by the dashed vertical lines. These lines mark roughly the extension of a (physiological) cortical column. Roerig et al. \cite{Roerig02} also noted that 'A small fraction of inputs originated more than one $mm$ away'.
While the displayed data might suggest two power laws, across the columnar scale we will work with an exponentially decaying connectivity probability. The difference to a power-law decay would here be small, and the exponential decay allows the direct comparison of our results to similar work performed in Ref.~\cite{Haeusler07}. An extremely broad model survey will demonstrate that inner-columnar wiring following biological templates have no computational advantage. At inter-columnar scales, i.e. networks taking whole columns as the nodes, a power-law decay will be implemented, and be compared to other connectivities. Power law connectivity will be found to be most beneficial if beyond the data range covered by Roerig et al. this power-law decay gives way to a milder power-law, expressing the observation that over very long distances, the connection probability should go to zero not too quickly. Motivated by an approximate self-similarity over the microcolumn-column-hypercolumn scales, we will assume that the exponent associated with the slow decay will be close to the one estimated across the columnar distance. Our results do, however, not critically depend on the exactly values of the exponents, only on their relative ordering is of relevance. Our implementation leads to a network that optimizes information transfer at minimal wiring cost.
 \begin{figure}[h!!!!!!!!!!!!!!!]
\begin{center}
\includegraphics[width=15cm]{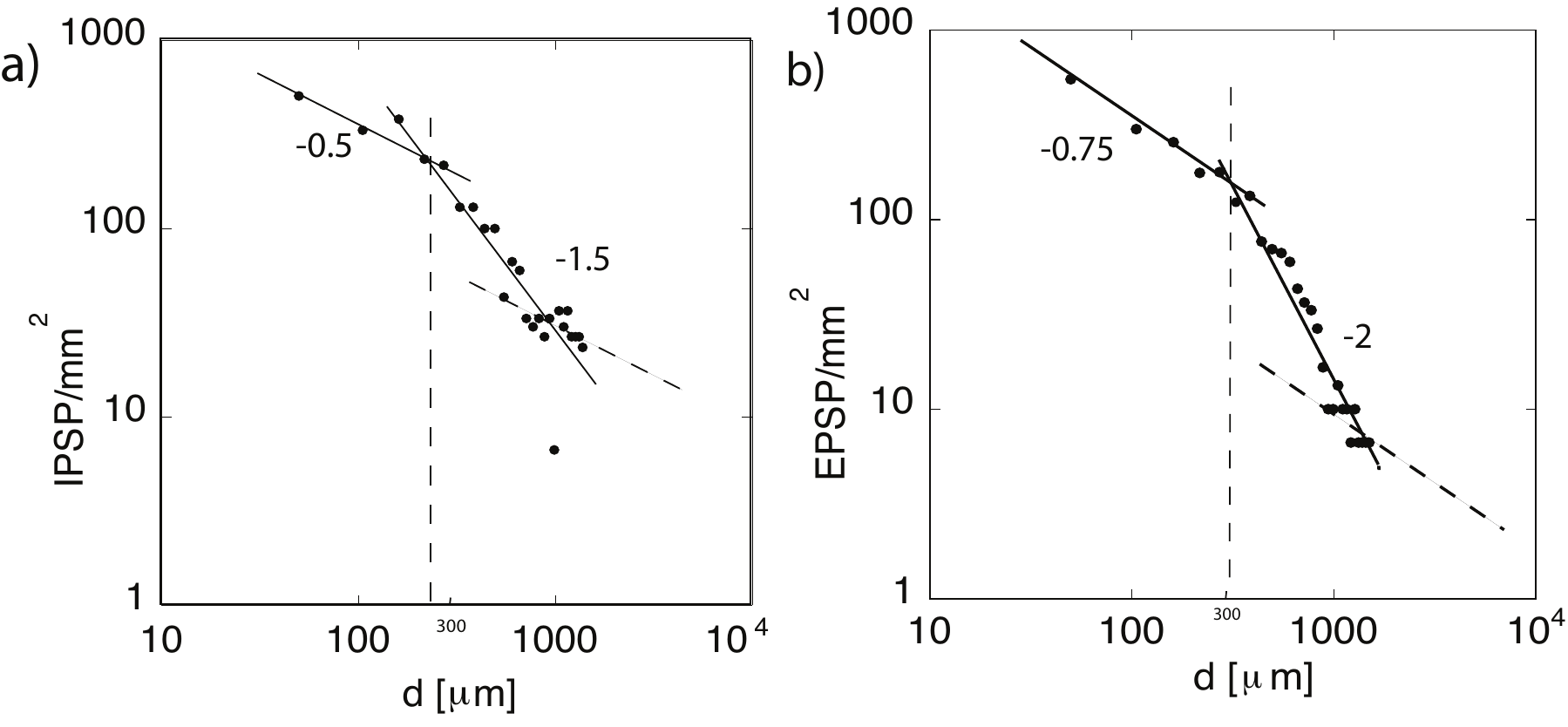}
\end{center}
\caption{\label{fig4a}\small{Biological data: Logarithmic density of photostimulation-evoked
excitatory (a)  / inhibitory (b) synaptic inputs %/ mm$^2$, 
in concentric rings spaced 50 $\mu$m apart, from 19 pooled layer 2/3 neurons (adapted from %Fig. 3AB in 
Ref.~\cite{Roerig02}). A small fraction of inputs originated more than one mm away. The vertical dashed lines mark the typical extension of a (physiological) column. While the data might suggest the presence of two power-laws, we will computationally implement power-law connectivity only across the intermediate scale. On the columnar scale, we will work with an exponential decay. Beyond the scales represented by the data, we propose the existence of a power-law of slower decay (tilted dashed lines) ensuring the existence of a minimal amount of connections across large scales.}} 
\end{figure}

{\bf Columnar computational nodes:} Despite one hundred years of intense investigations a precise correspondence between functional and physiological columns has not been obtained \cite{Horton}. Nonetheless, the wide-spread conception is that cortical computation takes essentially place within a column. To test the effect of wiring structure on computation, we therefore first measure to what extent inner-columnar wiring contributes towards computation. We implemented two levels of cortex-inspired wiring structure (cf. Fig.~2). Upon gradually eliminating architectural details by randomly rewiring the connections, we will measure the impacts that connectivity details have on cognition and computation. For a network realization, only connection {\it probabilities} shall be prescribed, contrary to what the term 'cortical microcircuit' used in \cite{Haeusler07} might evoke. Throughout all experiments, the abundance of inhibitory neurons within the population of neurons was kept at 20 percent.

\begin{figure}[h]
\begin{center}
\includegraphics[width=12cm]{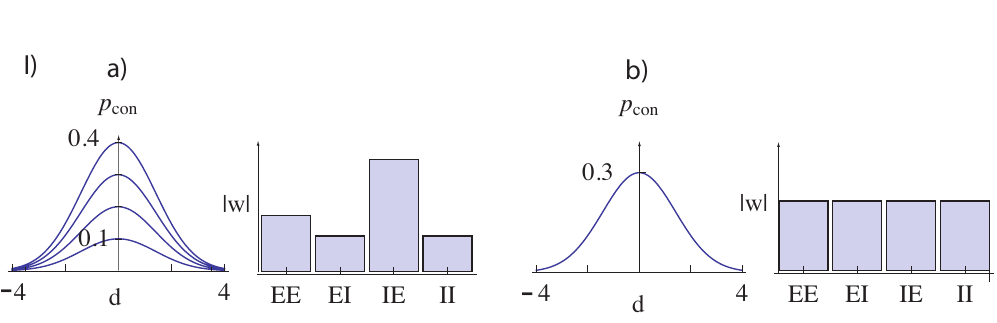}\\
\includegraphics[width=12cm]{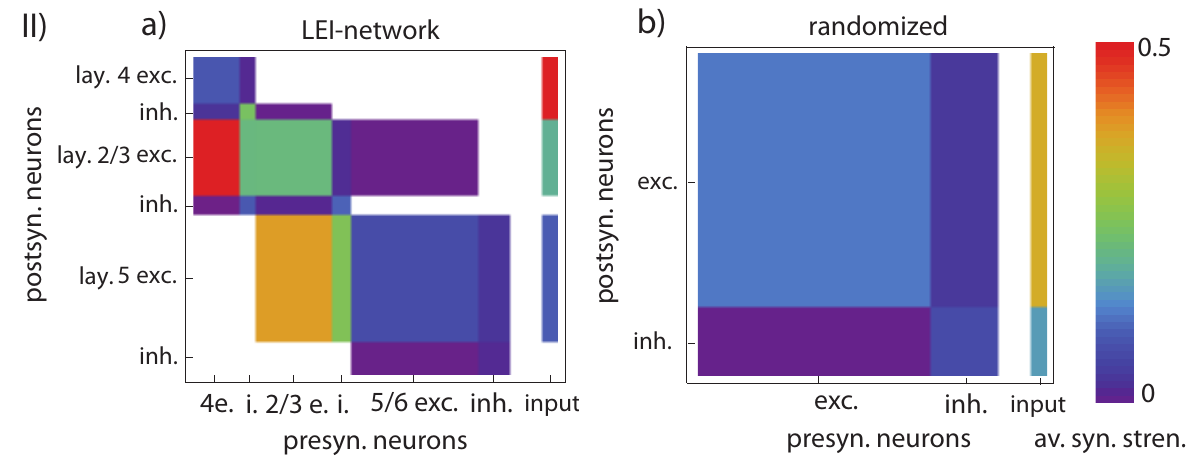}
\end{center}
\caption{\small{I) a) EI-model, b) EI-control network  (uniform synaptic weights $w$, $\lambda=~2$). 
$p_{con}$: probability of a synaptic connection among neurons of distance $d$ for $C$-values as in the text, $w$: synaptic strength of the connections. II) a) LEI-model, b) LEI-control network. The input vectors displayed at the right hand side of the matrices show how much input the respective populations receive.}}
\label{fig11}
\end{figure}
In the simple excitatory-inhibitory EI network model, the biological architecture is reduced to an excitatory and an inhibitory neuronal population and the connections within and between them. To vary the network structure within the EI model frame, we use a parameter $\lambda$ ruling the probability for a connection from neuron $j$ to neuron $i$ according to
\begin{equation}
p_{con}(i,j)=C(i,j)\cdot \exp\left(-\frac{d_{i,j}^2}{\lambda^2}\right),
\end{equation}
where $$d_{i,j}=|\hat{\mathbf{x}}_i-\hat{\mathbf{x}}_j|$$ is the Euclidean distance between the $i$'th and the $j$'th neurons' positions in the neural network (see end of paragraph). As $\lambda$ controls both the number and the typical length of the connections, varying from unconnectedness ($\lambda=0$) over local next-neighbor connectivity ($\lambda=1$) to global connectivity ($\lambda=\infty$), this parameter we will use to scan different network structures. $C(i, j)$ establishes the connectivity among excitatory (='E') and  inhibitory (='I') neurons, established by means of one pooled synapse. Our choice $C(E,E)=0.3$, $C(E,I)=0.4$, $C(I,E)=0.2$ and $C(I,I)=0.1$ reflects the typical biological connectivity. If a connection is made, the synaptic weights are drawn from a uniform distribution over $[0,1]$, multiplied by the weight factors $w(E,E)=30$, $w(E,I)=-19$, $w(I,E)=60$ and $w(I,I)=-19$.
%Rather than being geometrically segregated into layers, in this model the neurons are tagged with different properties. 
The model is compared to a control network where $C$ is uniformly set to $0.3$, the synaptic weights are again drawn from a uniform distribution over $[0,1]$, but endowed with a uniform weight factor scaled to match the total weight of the non-control networks and endowed with a sign to distinguish between inhibition and excitation. With the help of $\lambda$ we can assess a huge range of neural network architectures. There is, however, one important issue that we need to take care of additionally. Whereas in classical neural network theory the role of hidden neurons is crucial, in the classical liquid state network paradigm that we will use below, the input is relayed to every reservoir neuron, which renders all neurons non-hidden. To compensate for this shortcoming, we will also vary the fraction $I$ of input-receiving reservoir neurons. To simplify the comparison of results, we will mainly discuss results obtained for EI networks chosen as in Ref.~\cite{Maass02} on a three-dimensional grid of $3 \times 3  \times 15 = 135$ neurons. We checked, however, that the obtained results also hold for larger networks and exhibited, where this was not the case. 

%\subsubsection{Layered excitatory-inhibitory network topology (LEI)}
In the more detailed LEI network model, also the biological layering structure is implemented.
The LEI network is composed from three layers (2/3, 4 and 5/6), each of them containing an excitatory and an inhibitory population. The network contains again 135 neurons, with recurrent connections within the individual layers and connection probabilities and strengths as in Ref. \cite{Haeusler07}. As in the biological example input mostly feeds into layer 4 (input stream 1 in \cite{Haeusler07}). Layer 2/3 is the hidden layer, the output neurons are confined to layer 5/6. A family of control networks parametrized by $p \in[0,1]$ is obtained by replacing at each synapsewith probability $p$ the pre- and postsynaptic neurons by neurons chosen from the pooled neuronal ensembles of the same kind (excitatory or inhibitory). This rewiring procedure retains the overall connectivity and weight distribution between the excitatory and inhibitory populations, but gradually removes the three-layered structure. 
%This allows us to examine the impact of the segregation of LSN into layers and interlayer connectivity.
%\begin{figure}[h]
%\begin{center}
%\includegraphics[width=8.5cm]{spiketrains.eps}
%\end{center}
%\caption{LEI circuit experiment: a) Input stimulus (Arabic Digit MFCC), b) evoked spike trains. 
%In the paradigm excitatory and inhibitory neurons remain at their positions.} 
%\end{figure}
%Together with the amount of input flowing into the network, the synaptic weights determine the average activity of the neurons. To arrive at comparable levels of synaptic transmission for all circuits, $\mathbf{W}$ was scaled to a common largest eigenvalue (1 in EI networks, 0.2 in LEI networks). Our choice of parameters ensures that without reliance on input, neurons can be excited by their presynaptic partners.
%For all experiments performed it was verified that with our choice of parameters (and while the neurons were excited from both input and presynaptic spikes), the firing rates but occasionally vanished or saturated (where $f_{sat}=\frac{1}{\tau}$). This ensured that during the experiments, the network's activity was kept within a favorable dynamic range for all network topologies.
%In Fig.~2, we exhibit the temporal output of the output neurons of the LEI network and its control, upon Arabic Digit recognition learning. From the output spike patterns, a superiority of the biologically detailed reservoir cannot be inferred.

The measurement of the computational effect of inner-columnar wiring structure is done within the framework of a reservoir computing neural network. While not in all aspects of top-class efficiency among the possible network types \cite{Ganguli}, reservoir computing has successfully been used in robot motion planning \cite{Jaeger07}, despite the linear decision boundaries that it implements. In these networks, learning is confined to the network's periphery (see Fig.~3 for a conceptual drawing), which allows to assess the pure effect of the inner-columnar wiring structure on computation without being compromised by the learning process. In the original versions of reservoir computing, the neurons of the 'reservoir' are randomly connected in a recurrent fashion. Reservoir neurons receive external input from the signal and recurrent input from other reservoir neurons. In order to parallel the biological example, we will implement models of spiking neurons, which turns the network into what is often called a Liquid States network \cite{Maass02, Maass03, Bertschinger04, Haeusler07}.  %It has already been remarked that Echo State Machines (ESN) with analog neurons and sigmoidal transfer functions show little performance sensitivity with respect to changes of the reservoir topology \cite{Jaeger07,Lukosevicius09}. In the quest of understanding the brain, following the idea to take the recurrently connected cortex as the blueprint for artificial neural networks, spiking neurons were introduced into the reservoir. This led to the "Liquid State Machines" (LSM) network paradigm (\cite{Maass02}). In these networks, a marked performance increase was reported if within their reservoir a refining topology was implemented that in some sense was 'close' to that found in the mammalian cortex \cite{Maass02, Maass03, Bertschinger04, Haeusler07}. This interesting phenomenon was investigated in details in \cite{Haeusler07}, by comparing an LSM endowed with a layering connectivity as dealt with in cortex, with LSM networks based on random wiring. For the layered network, a significantly improved computational performance was reported. Recent attempts \cite{Schrauwen08,Buesing10} to resolve the apparent contradiction between ESN and LSM performance attributed the phenomenon to the difference between the hyperbolic tangent vs. the Heaviside transfer function, i.e. to analog vs. digital/spiking neuronal information processing used in these paradigms. One finding was that LSM built on spiking neurons with digital membrane potential, were superior to fully analog neuron based LSM, particularly in conditions of sparse connectivity (typically 3 synapses/neuron) or sparse activity (typical for biological conditions), the reason being that the former more easily reach the region of optimal synaptic input. This argument, however, fails in the case of LSM built on spiking neurons with a continuous membrane potential. 
%In these investigations, to what extent the reported differences would be measurable or instrumental in real world applications, was only vaguely dealt with. Answering this question is main topic that we will be concerned with here. In the cortex, an abundant number of neuron geometries and of spatial and functional connectivity units (so-called cortical layers) have been identified \cite{Binzegger04}. For our study it was necessary to limit the amount of incorporated details. The modeling of biologically layered circuits ('columnar neural organization') as performed in our study will, however, reflect the correct population ratios of the involved neuronal ensembles, interlayer connection probabilities, and synaptic weight distributions as encountered in the cortex. 
Ref. \cite{Haeusler07} reported a marked performance increase when columnar fine-structure was added into Liquid States networks. Their evidence was,  however, extracted from artificial contexts that were not tested for practical relevance. We will first show that for real-world cognitive tasks, we do not find signatures of such a computational boost. Then we will zoom out from the columnar to the inter-columnar scale, which will offer an alternative argument for the existence of cortical columns.
%As the general result, we fail to see the expected  benefits.  %Also from running unmodified packages from Ref.\cite{Haeusler07} (where they made their simulation software publicly available), we did not arrive at an equally optimistic view. 
%\section{Methods}
%\subsection{The Liquid State Machine Concept}
\begin{figure}[h] 
\setlength{\unitlength}{0.15in}   % selecting unit length {0.097in}

\centering       % used for centering Figure 
\begin{picture}(35,10)   % picture environment with the size (dimensions) 
      % 32 length units wide, and 15 units high. 
\put(1.5,4){\framebox(4,3){$\mathbf{W}_{in}$}} 
\put(6.75,-0.5){\dashbox{.2}(12,10){}}
%\put(10.5,10) {Reservoir}
\put(10.4,8.5) {Reservoir}
\put(7,4){\framebox(5,3){Neurons}} 
\put(13.5,4){\framebox(5,3){Synapses}} 
\put(11,0){\framebox(4,3){$\mathbf{W}$}} 
\put(20,4){\framebox(5,3){Readout}}
\put(26.5,4){\framebox(4,3){$\mathbf{W}_{out}$}}
\put(0,5.5){\vector(1,0){1.5}}
\put(5.5,5.5){\vector(1,0){1.5}} 
\put(12,5.5){\vector(1,0){1.5}}
\put(16,4){\line(0,-1){2.5}}
\put(16,1.5){\vector(-1,0){1}}
\put(11,1.5){\line(-1,0){1}}
\put(10,1.5){\vector(0,1){2.5}}
\put(18.5,5.5){\vector(1,0){1.5}} 
\put(25,5.5){\vector(1,0){1.5}} 
\put(30.5,5.5){\vector(1,0){1.5}} 
%\put(28.5,2){\vector(0,1){2}} 
\put(10,7){\line(0,1){1}}
\put(10,8){\line(-1,0){1}}
\put(9,8){\vector(0,-1){1}}
%\put(9.2,8.5){a)}
\put(16.5,7){\line(0,1){1}}
\put(16.5,8){\line(-1,0){1}}
\put(15.5,8){\vector(0,-1){1}}
%\put(15.7,8.5){b)}
\put(23,7){\line(0,1){1}}
\put(23,8){\line(-1,0){1}}
\put(22,8){\vector(0,-1){1}}
%\put(22.2,8.5){c)}
\put(-1,5.2) {$\mathbf{u}$}
\put(12.3,6) {$\mathbf{v}$}
\put(25.4,6) {$\mathbf{x}$}
\put(19,6) {$\mathbf{s}$} 
\put(16.5,2.5) {$\mathbf{s}$} 
\put(32.5,5.2) {$\mathbf{y}$} 
%\put(28,1) {$\mathbf{y}_d$} 
\end{picture} 
\caption{\small{ Reservoir computation model: Stimulus ${\bf u}$ is associated with output ${\bf y}$. Input ${\bf u}$ is relayed via weight matrix $\mathbf{W}_{in}$ to the reservoir. In the drawing, the synapses in the reservoir (${\bf s}$) are artificially separated from neuron membrane potential (${\bf v}$) dynamics and the readout neurons are artificially separated from the rest of the reservoir. Recurrent reservoir topology is encoded in matrix $\mathbf{W}$ via the choices of ${\bf s}$. 'Learning' is confined to matrix $\mathbf{W}_{out}$.\label{fig6} %Main information flow in the cortex is from layer 4 (input amplification)$\rightarrow$layers 2/3$\rightarrow$layers 5/6$\rightarrow$layer 4 (activity stop).
}} 
\end{figure}
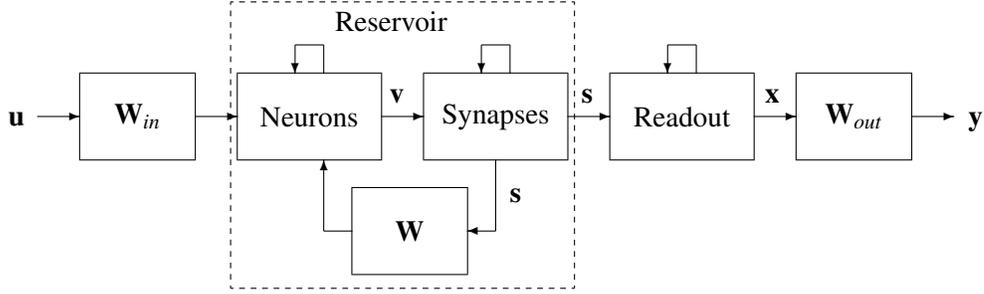
%Starting from the cortical blueprint, we can probe the architecture's influence on cognition and computation by gradually randomly rewiring the neuronal connectivity and measuring the obtained performance.

Reservoir computing is a supervised process to associate $k$ pairs $$\{\mathbf{u}(t)_i \,,\, \mathbf{y}(t)_i\}_{i \in\{1,..,k\}}$$ of input $/$ output sequences of individual sequence length $T_i$ (so that $ t\in\{1,..,T_i\}$). The dimensionality of the input vectors is denoted by $N_u$, the dimensionality of the output vectors  by $N_y$. Upon stimulation by the input sequence $\mathbf{u}(t)_i $, the liquid reservoir of $N_x$ neurons generates a state vector $\mathbf{x}(t)_i$ of the same dimension. Let $T=\sum T_i$ denote the total time spanned by the input patterns, and let  $\mathbf{X}$ be the $N_x \times T$-matrix of states. Let $\mathbf{Y}$ denote the $N_y\times T$-matrix of the associated patterns.
The desired relation $\mathbf{W}_{out}\mathbf{x}(t)_i\simeq{\mathbf{y}}_d(t) _i$ leads directly to the 
least-squares optimized read-out matrix $$\mathbf{W}_{out}\simeq \mathbf{Y} \mathbf{{X}^+},$$
if $\mathbf{{X}^+}$ is the (Moore-Penrose) pseudo-inverse of $\mathbf{X}$. 
%As the implementation used for our investigations, we used the pseudo-inverse implementation offered by the program packet Ref.\cite{pseudoinverse}.
In typical applications, the system should respond to a stimulus $\mathbf{u}(t)$ by the desired temporal pattern $\mathbf{y}(t)$. During the learning phase, the input and the desired output signal are fed into the reservoir. Care is taken that the scaled input signal optimally stimulates the respective target neuron models (see below). After a transient phase, the optimized output matrix $\mathbf{W}_{out}$ is calculated. This step corresponds to the learning process in classical neural networks.
In this framework, network realizations are captured in the connection weight matrix $\mathbf{W}$, whereas 'learning' in the traditional sense is confined to the read-out matrix $\mathbf{W}_{out}$. Excitatory connections are reflected by positive synaptic weights ${\bf s}$, inhibitory connections by negative weights and absence of connections by zero weights.

For all networks, the average neuronal activity is determined by the overall scaling of the input and the synaptic weights (and in the case of the Izhikevich neurons also by the background current) and by the wiring matrix $W$. To ensure that networks of a similar level of neuronal activity are compared, the matrices $W$ chosen according to the wiring model were scaled to obtain a common largest eigenvalue (1 for EI networks, 0.2 for LEI networks). Matrix $W_{in}$ was chosen by drawing from across $[-0.2,\,0.2]$ uniformly distributed random numbers (in the case of Izhikevich neurons (see below) scaled by a factor of 30, to arrive at the standard parameter scale). The input and  synaptic efficiencies were scaled so that neurons could be excited by their presynaptic partners without reliance on input, but that the firing rates in response to excitation from both input and presynaptic spikes were also sufficiently distant to saturation ($f_{sat} = 1/\tau$). This procedure and the chosen parameters ensured that during all the parameter sweeps we performed, the network was confined to the same dynamic range. %As the time step of the input data (10 $ms$) was larger than the time step of the neural simulation ( = 2 $ms$), across the intermediate time steps the input was held constant.

 %obtained as the pseudo-inverse of the matrix of dimension $T \times M$ applied to the desired output vector $y_d(t)$ of dimension $T$ (where $M$ is the dimension of the output vector, often equal to $T$). The optimization process is embodied by taking the pseudo-inverse of the $T \times M$-dimensional matrix $\mathbf{{X}^+}$, applied to the matrix of dimension $k \times T$ of the desired outputs $\mathbf{Y_d}$, where $k$ is the cardinality of the patterns. In the case of associative tasks, often a minimal distance classification is performed.
%\subsection{Neuron models}
%\subsection{Cortical Microcircuit Topologies}

We used two methods of reading out the reservoir. In the original Liquid States network, the readout is immediate, i.e. memoryless ($'ml'$). For every input vector an output vector is generated, allowing "anytime recognition" \cite{Maass02}. For classification tasks it may be advantageous to have a memory span of a size comparable to the stimulus length. Otherwise the Liquid States network will confuse stimuli containing similar parts (e.g. phonemes in speech recognition).
This is why we also used an alternative read-out method. For every neuron and stimulus, a stimulus-averaged firing rate yielded the read-out vector. This method (referred to as integration ($'int'$) readout) leads to significantly improved recognition rates, although it deviates from the usual Liquid States network paradigm. 

\begin{figure}[h]
\begin{center}
\includegraphics[width=15cm]{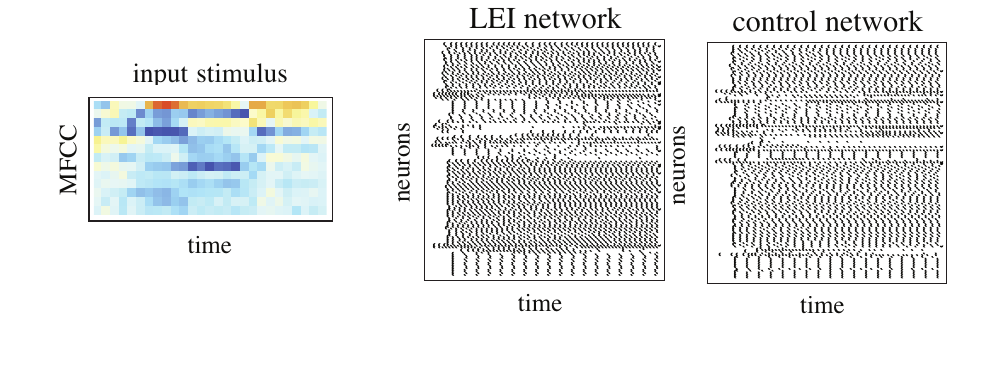}
\end{center}
\caption{\small{Arabic Digit MFCC input (color-coded), processed by a LEI neural network with spiking neurons and by its control network. The spike trains of the two networks are similar because the position number of the excitatory and of the inhibitory neurons are maintained. }}
\label{fig5}
\end{figure}

To arrive at statements that are largely independent of the network elements, we tested two rather distinct models of the neuronal membrane voltage dynamics  $v(t)$ in Fig.~\ref{fig6}. The leaky integrate-and-fire neuron dynamics is defined by
$$v_i(t+\tau) = \exp(-\frac{\tau}{\tau_m}) \cdot v_i(t)+\sum_{j}^{N_x} w_{ij} s_j(t)+\sum_{l}^{N_u} w_{in_{il}} u_l(t).$$ $s_j$ is the postsynaptic potential at the synapses innervated by the $j$'th neuron, which is weighted by the synaptic efficiency $w_{ij}$ between presynaptic neuron $j$ and postsynaptic neuron $i$.  $u_l$ denotes the $l$'th input component which is weighted by the connection strength $w_{in_{il}}$ from the $l$'th input component to neuron $i$. We use a membrane time constant $\tau_m=30$ ms, $\tau$ is the integration time-step. If $v_i$ reaches the threshold $V_{thr}=1$, a spike is triggered, which resets the synapse $s_i$ to 1 and $v_i$ to $v_i=V_{res}=0$.
The fast-spiking simple Izhikevich neuron dynamics \cite{Izhikevich03} is given by 
the coupled equations 
\begin{equation*}%\scriptsize{
v_i(t+\tau) = v_i(t) + \tau \left[ \frac{4}{100}\,v_i^2(t)+5\,v_i(t)+30\left(\sum_{j}^{N_x} w_{ij} s_j(t)+\sum_{l}^{N_u} w_{in_{il}} u_l(t)\right) - r_i(t) + 152\right],
%}
\end{equation*}
\begin{equation*}%\scriptsize{
r_i(t+\tau) = r_i(t) + \tau \left[ \frac{2}{100}v_i(t+\tau)-\frac{1}{10}\,r_i(t)\right].
%}
\end{equation*}
The additional variable $r_i$ controls sub-threshold dynamics and refractoriness. If $v_i$ reaches the threshold $V_{thr}=30$, a spike is triggered, which resets $s_i$ to 1, $v_i$ to $-65$ and $r_i$ to $r_i+2$.
For both neuron models, the synaptic signal experiences an exponential decay
$s_i(t+\tau)=\exp\left(-\frac{\tau}{\tau_{Syn}}\right)s_i(t)$,
with a constant $\tau_{Syn}=2$ ms. The axonal delays and the refractory periods are determined by a fixed integration step of $\tau =$2~ms. 
The transcription from the dynamic synapses used in \cite{Haeusler07} to our exponential synapses is achieved by setting our synaptic weights equal to the steady state strength $U$ of \cite{Haeusler07}. %U is given as the strength of a dynamical synapse in the limit of vanishing facililitation and depression times.
An example of an input signal processed by reservoir networks is given in Fig.~\ref{fig5}.

%To conform with the situation encountered in the cortex, the ratio of inhibitory to excitatory neuronal population sizes was chosen to be 20\%.
%For each experiment, the elements of matrix $\mathbf{W}_{in}=(w_{in})_{ik}$ are independently drawn from the uniform distribution across the interval $[-0.2,0.2]$. Since the time step of the input data ($10$~ms) is larger than the time step of the neural simulation ($\tau = 2$~ms), the inputs need to be interpolated during the intermediate steps. This is achieved by a constant input value in this period.

{\bf Networks of columns:} When zooming out from the columnar dimension to the inter-columnar scale, we have to wrap up the columnar computation and relate it to the computations performed by other columns. Rulkov \cite{Rulkov1,Rulkov2} has demonstrated that any desired neuronal firing behavior representing columnar response can be expressed by a suitably chosen discrete map. The natural model then to use is a coupled map lattice  \cite{Kaneko,Bunimovich} of chaotic maps, to have the response flexibility required for communication. In our modeling of bio-inspired connectivity, the probability that two lattice sites $i,j$ of distance $d_{i,j}$ are connected is 
\begin{equation}p_{i,j} = \theta \cdot {d_{i,j}}^{-\alpha} + (1-\theta) \cdot {d_{i,j}}^{-\beta},\end{equation} which specifies the connectivity matrix, see Fig.~\ref{fig4b}a. By choosing $\alpha$, $\beta$ and $\theta$, a range of network architectures similar to those explored in the Liquid States network paradigm can be accessed. Given $\theta = 1$, the system can be changed from a globally coupled network ($\alpha = 0$) into a nearest-neighbor coupled network  ($\alpha \rightarrow \infty$). For $0 < \theta < 1$, $\beta = 0$, $\alpha \rightarrow \infty$, the network is coupled to the nearest neighbor with probability 1 and to all other nodes with probability $(1 - \theta)$ up to the cutoff $M$. As a result we obtain a combined nearest neighbor- and random-coupled network. For all intermediate values of $\alpha$ and $\beta$, the network is fractally coupled. The cutoff value $M$ determines, together with the underlying topology, the average number of connected nodes $k$. 
The interaction of the local chaotic site maps $f$ is modeled by diffusively coupled evolution of the site states 
\begin{equation}x_i(t + 1) = (1-\varepsilon)f(x_i(t)) +\frac{\varepsilon}{A_i}\sum_{j\in conn} f(x_j(t)),\end{equation}
where $t$ denotes discrete time, $A_i$ the number of connections to the $i$'th site indexed by $j$, and $\varepsilon$ the coupling strength. 

The computation performed at this scale will be characterized by the network's ability to quickly propagate incoming information and by its ability to generate among the affected cells a coherent state, expressing that 'computation' has emerged \cite{Stoop}. Both abilities depend crucially on the number of connections $k$ that impinge on a cell. 
Spatio-temporal information propagation is the maximal velocity of the propagation of perturbations through the network. %: A small perturbation applied to the oscillators of a columnar cluster,
%and the information propagation is followed using the difference to a replica system without
%perturbation. The information propagation velocity $v^{*}$ can directly be measured from the
%perturbation at the leftmost and the rightmost oscillator. 
As a basic approximation, this propagation is the result of two independent contributions: The chaotic
instability of the map leads to an average exponential growth of the initial infinitesimal perturbation $d_0$ applied at site $0$,
whereas the diffusive coupling that results in a Gaussian spreading. The combined perturbative effects at site $i$ are
then expressed by the equation  \cite{Cencini}
$$|\delta x_i(t)| \approx \frac{d_0}{\sqrt{4 \pi D t}} e^{\tilde{\lambda} t-\frac{i^2}{4 D t}},$$
where $D$ denotes the diffusion coefficient, $\tilde{\lambda}$ is the Lyapunov exponent of the site map,
and $d_0$ the perturbation strength. The velocity $v$ of the traveling wave front is determined at the
borderline of damped and undamped perturbations, which implies
that $v$ depends on $D$ according to  \cite{Cencini} \begin{equation}v = 2\sqrt{\tilde{\lambda}}\cdot \sqrt{D}.\end{equation} 
For a given site map $f$, the speed of information transfer ('SIT') is therefore determined by $\sqrt{D}$ and 
all that remains to be done is to estimate $D$ from the network via the mean transition time. 
This quantity can be determined from the network topology alone, using a Markov chain approach. %The expression shows that
%the particularities of the node maps may be left unspecified; only the degree of chaoticity of the network nodes is of interest (if the node maps are chosen to be identical, otherwise 'chaoticity' refers to a suitably defined ensemble average). 
Additional network features can be implemented via the connectivity matrix. Implementation of a detailed columnar 
structure left our total wiring length results unchanged.

Technically, the ability to generate a coherent state is expressed by the cells' ability to synchronize in a generalized sense. 
For synchronization, a minimal number $k$ of connections are required to impinge on a given site. Full dynamical synchronization of the chaotic sites
continues to exist if the condition  \cite{Atay} $\mid e^{\tilde{\lambda}}-\varepsilon \mu_k\mid <1$ (where $\tilde{\lambda}$ is the site map Lyapunov exponent and $\mu_k$ are the nonzero Eigenvalues of the graph Laplacian) is maintained. This simple criterion may be overly severe, but it is indicative of what will be found on finer levels of description as well. 

Different network architectures should therefore be compared under the constraint of an equal number of connections $k$.
Biologically relevant indicators for the efficiency of the network will then be the speed of information transfer through the network 
and the total wiring length required for synchronized columns.

\section*{RESULTS}
\begin{figure}[h]
\begin{center}
\includegraphics[width=9cm]{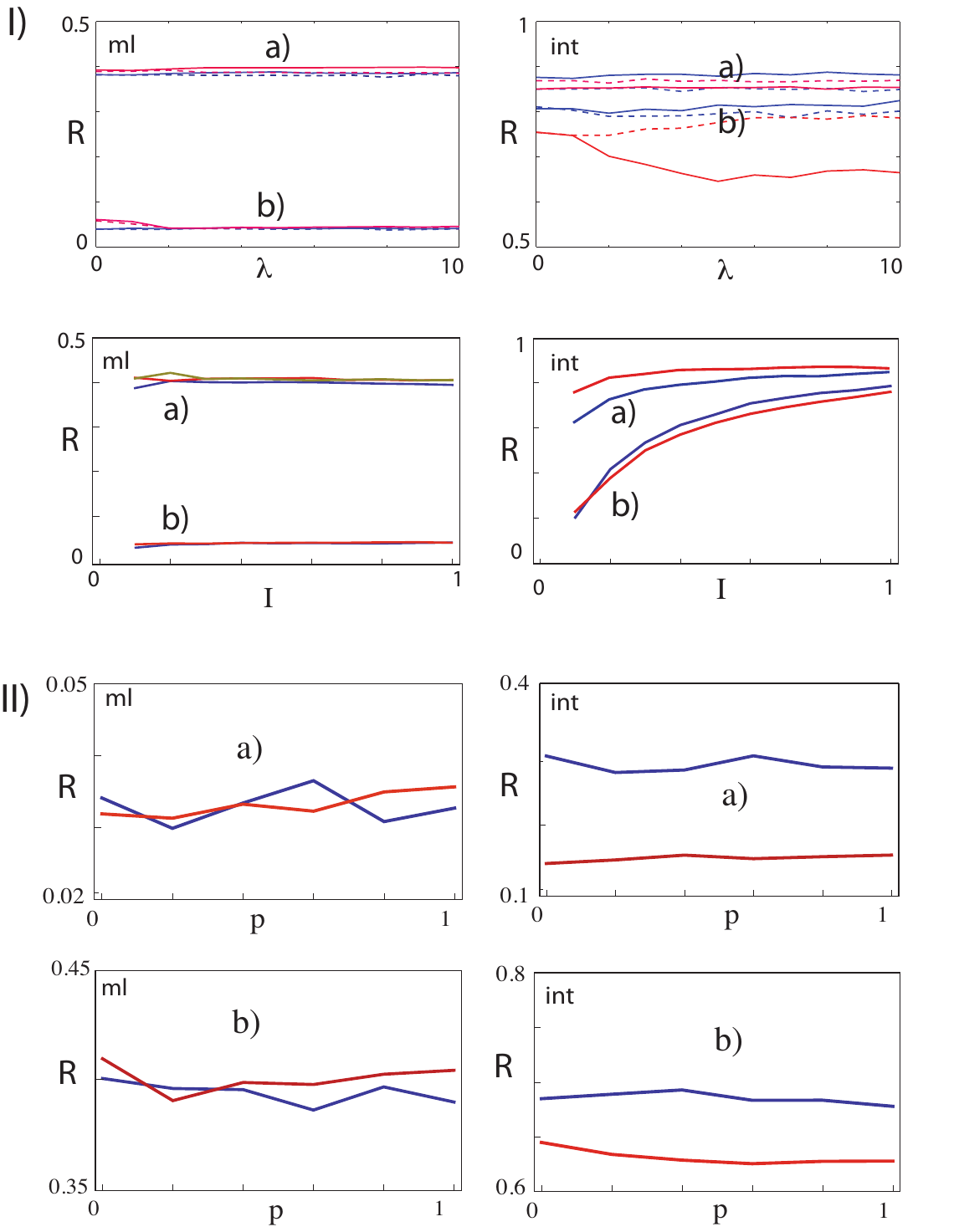}
\end{center}
\caption{\small{Recognition rate $R$ for a) Arabic Digit,  b) Auslan Sign recognition, using leaky integrate and fire (blue curves), or Izhikevich (red curves) neurons in the networks. Each data point is the average over 20 realizations. Left column: memoryless ('ml'), right column: integration ('int') readout. Networks (cf. Fig.~\ref{fig11}):
I) EI network, recognition rate dependence on connectivity $\lambda$ (control networks: dashed curves), and on the ratio $I$ of input receiving neurons at local connectivity at $\lambda=2$. Ocher: Izhikevich neurons with $\lambda=0$ (no connections). 
II) LEI networks, recognition rate dependence on the rewiring probability $p$. $p=0$: layered, $p=1$: homogeneous control network. }}
\label{fig1}
\end{figure}

On the columnar scale, where we focus on computation, two popular time series classification tasks serve as real-world benchmarks, in contrast with the more abstract computations considered in Ref.~\cite{Haeusler07}. Single Arabic Digit speech recognition \cite{Hammami09} is based on time series of 13 Mel Frequency Cepstral Coefficients for 10 classes of digits spoken by 88 subjects (cf. Fig.~4). Australian Sign Language (Auslan) recognition is based on time series of 22 parameters for 95 signs, recorded from a native signer using digital gloves and position tracker equipments \cite{Kadous02}. We investigated the influence of cortical organizational structures on two levels of architectural sophistication: A simpler excitatory-inhibitory EI network and the more detailed layered excitatory-inhibitory network topology LEI.
We start the discussion of our results from  Liquid States network with two general observations clear from Fig.~\ref{fig1}. Whereas the particular neuron models (and the underlying circuit parameters) are of secondary influence (blue vs. red curves), the integration readout (right panels) has a clear advantage over instantaneous readout (left panels).%The datasets are freely available \cite{UCIRep}. 
The results obtained for the EI-network demonstrate that the connectivity expressed in terms of $\lambda$ does not enhance the computational power of the network. The plot also confirms that local connectivity $\lambda \approx 2$ plays no distinguishable role among the possible connectivities.
Having no recurrent connections at all among the reservoir neurons does not hamper the recognition rate, suggesting that extremely little computation is owed to synaptic interaction \cite{Fette05}. %For Izhikevich neurons, reservoir connections can even result in a performance decrease. 
One may suspect that the low recognition rates from memoryless readout are because in the classical  Liquid States network paradigm the input signal is applied to all neurons, which constantly overwrites memory that otherwise would be retained in 'hidden' neurons. To exclude this possibility, we examined in Fig.~\ref{fig1}'s second row the role of the hidden neurons, by measuring $R$ for networks having a fraction $I$ of input signal receiving neurons. The desired value of $I$ is achieved by removing from a corresponding number of reservoir neurons the input signals. To compare with Refs. \cite{Verstraeten05,Maass02}, the connectivity was set to local next neighbors ($\lambda=2$), except for one test using Izhikievich neurons with $\lambda=0$ (ocher line).
%With both readout methods, we observe no benefit from hidden neurons.%; these neurons cannot remedy the low memoryless readout performance. 
If hidden neurons were beneficial, we, again, should perceive a maximum of $R$ for some optimal value of $I$. In the Arabic Digit task with memoryless readout we do not observe a dependence on the number of actually used neurons (i.e. beyond $I=0.1$, where we have on average $13.5$ input receiving neurons, at an input dimensionality of $13$). The similarity of the results obtained for $\lambda=0$ and for $\lambda=2$ suggests that the nonlinear interaction among the input receiving neurons does not significantly enhance performance. In the Auslan task we see a monotonous dependence of $R$ on $I$, because for most values of $I$ the number of input receiving neurons is smaller than the input dimensionality (i.e., we have $I \cdot135<95$). As a function of $\lambda$, in the biological setting Izhikievich neurons tend to globally phase-lock to the strong inner excitation, which leads to a somewhat reduced input-responsiveness. The details of why the effect emerges exactly in the biological setting is not clear.
The EI network with biology-motivated wiring structure thus does not perform significantly better than the control network. 
The results obtained for the LEI networks reflecting to more details the columnar layering structures (see Fig.~\ref{fig1} II) corroborate the observations made for the simpler model: A significant dependence of $R$ on the rewiring probability $p$ was not observed. %The layer segregation and the special interlayer connectivity of this circuit does not lead to an improvement over a monolith random circuit. 
%The recognition enhancement by the integration readout seems therefore to be an effect of input/output neuron sparseness. %\section{Discussion}
%In our experiments we focused on time series classification problems, because we believe this class of tasks to be typical for the everyday work performed by the mammalian cortex. 
%Across a wide variety of LSN network realizations, we consistently found no evidence of a micro-architectural performance boosting. 
These observations are compatible with earlier findings for echo state networks \cite{Jaeger07}. 

On the mesoscopic scale, instead of the recognition rate we focus on the information transport across the cortical network, and on the material (i.e., the network's total wiring length) needed for obtaining a spatio-temporally coherent computation. To demonstrate the beneficial effects by a columnar organization of the network, a coupled map lattice \cite{Kaneko,Bunimovich} is now a more appropriate network model than the Liquid States network (see last section). In the coupled map paradigm, neuronal interaction is modeled by chaotic site maps communicating by way of diffusive coupling (Eq. 3).
\begin{figure}[h!!!!!!!!!!!!!!!]
\begin{center}
\includegraphics[width=15cm]{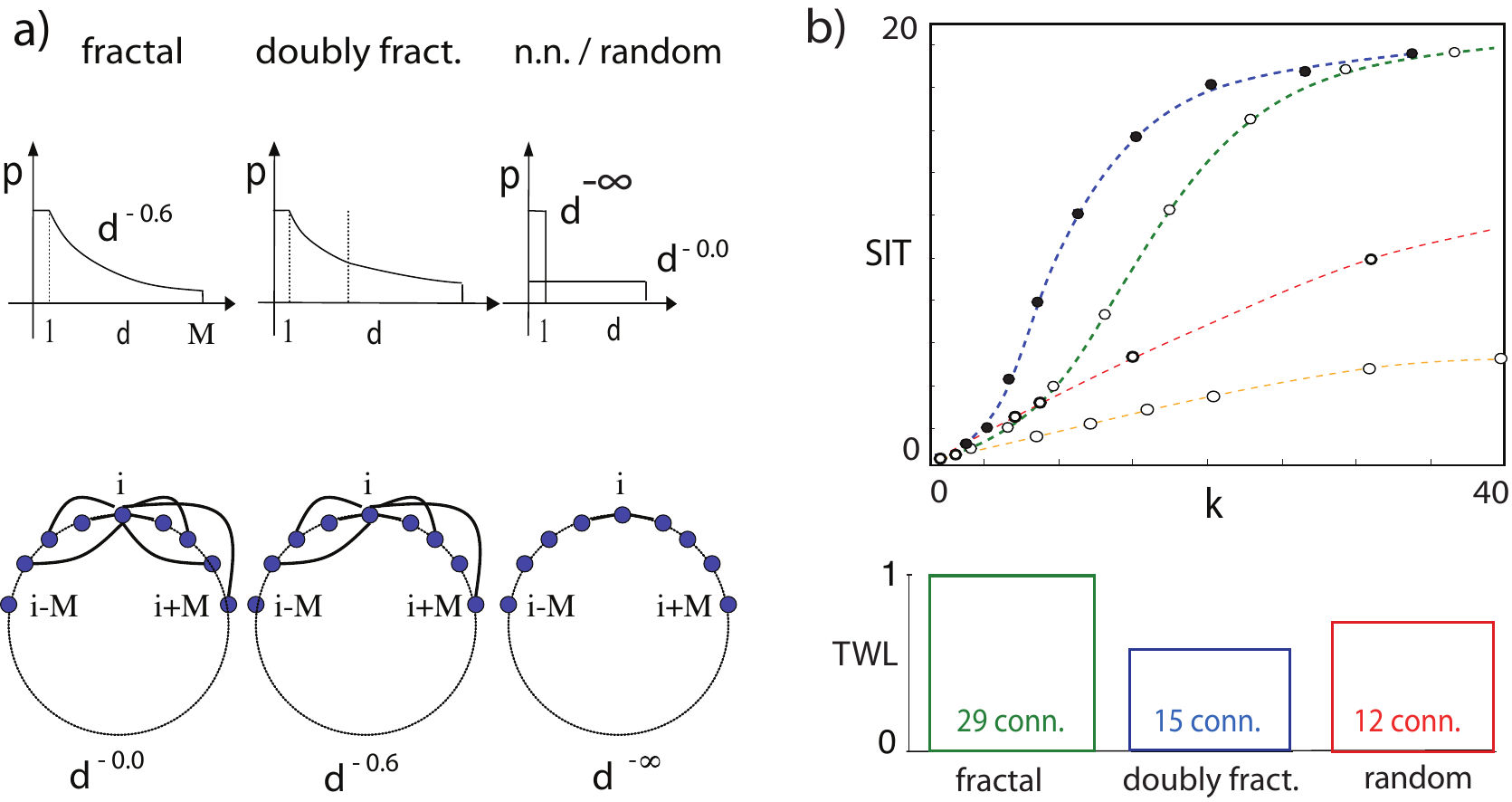}
\end{center}
\caption{\label{fig4b}\small{a) Main network densities compared (p: connection probabilities, d: distance, M: cutoff, see text). Below: examples of 'fractal' wirings. b) Speed of information transfer 'SIT' as a function of average number of connections $k$ established by Eq. 2. From top: doubly fractal ($\theta=0.2,\, \alpha=0.5, \, \beta=2.0$), fractal ($\theta=1, \, \alpha=0.7$), random, next-neighbor topology. Networks sizes: $N=4096$. Data points are averages over 100 realizations. Lower panel: Typical number of connections $k$ required for synchronization (numbers) and corresponding relative total wiring length TWL (histogram height). $N=512$. Average over 10 realizations.}}
\end{figure}
 When we measured the speed of information transport through the network for doubly fractal, single fractal, random, and nearest-neighbor, coupling topologies, we indeed found a strong dependence on the wiring topology. For the doubly fractal architecture suggested by the data of Roerig et al. we find a consistently enhanced speed of information transfer if compared to the alternative networks (Fig.~\ref{fig4b}b), upper panel). In this figure we plot the speed of information transfer in arbitrary units, that scale by the square-root of the positive Lyapunov exponent characterizing the chaotic site maps (Eq. 4). Speed enhancement persists across a wide selection of pairs of exponents as long as the qualitative size of the exponents is preserved \cite{StoopWagner}. %We suspect that a preciser cortex model should even be tri-fractal, where a milder decay for longer connections completes the above picture (cf. Fig.~\ref{fig4} a). Several observations, however, indicate that our findings also hold for this case.
Our numerical experiments show that under the condition of synchronization, the doubly fractal architecture jointly optimizes total wiring length TWL and speed (Fig.~\ref{fig4b}b), lower panel).%Details of how to determine the numbers of neighbors $k$ necessary for achieving the threshold to synchronization, are provided in Ref.~\cite{StoopWagner}. %fails to minimize the numbers of neighbors.

\section*{DISCUSSION}

Our inner-columnar wiring computational experiments focused on time series classification problems, because this task class seems to be typical for  everyday work performed by the mammalian cortex. 
Across a wide variety of network realizations, we found no evidence of a micro-architectural performance boost. A significant dependence of $R$ on the rewiring probability $p$ is in our experiments not detectable. Layer segregation and special inter-layer connectivities do not lead to improvements over fully random circuits. The recognition enhancement observed with the integration readout is an effect of input/output neuron sparseness.
Obviously, much more work on understanding the dynamics and computations in recurrent neural networks needs to be done if we would like to argue that observed detailed biological wiring schemes facilitate efficient computation. At the present stage, it rather seems that exact connection statistics play no eminent computational role. Instead, their role could be to keep the neurons in an overall dynamic regime, allowing them to be maximally input-sensitive. Computational benefits could still be located in the neuronal fine structures. This level is necessarily disregarded when modeling generic columns from measurements across a large number of cortical sites, subjects or even across species. 
Our focus on given biological data, and the particular comparisons we decided to deliver, did not allow us to fully exploit the power of liquid states networks: Higher recognition rates could be achieved with a larger number of neurons and with different time constants. In particular, in practical applications of liquid states networks, the amount of computational resources occupied by the synaptic interactions would more wisely be invested in an increased neuronal population. This would also have the advantage of rendering these networks parallelizable in a simpler way. These constraints do, however, not compromise the central finding made in this part of the work: The absence of the claimed computational boost by the biologically motivated wiring schemes.

To a real-world neural network, propagation speed and computation across the mesoscopic inter-columnar scale are of equal importance with local computation. At this scale, the network is consistently modeled as a coupled map lattice, with its columnar outputs captured by chaotic maps communicating via diffusive coupling.
We found that the few long-ranged connections present in our doubly-fractal model, substantially enhance the speed of information transfer beyond those provided by concurrent network topologies. Thus, the columnar structure may have emerged as a facilitating structure for speed of information transfer optimization, irrespective of the particular values of the exponents $\alpha, \, \beta$ characterizing the connectivity decay. Moreover, and more importantly, doubly fractal networks synchronize at a shorter total wiring length TWL and, at this condition, keep their superior speed of information transfer SIT. 
The doubly fractal networks achieve this performance irrespective of whether the long/short ranges are implemented by neurons with long {\it and} short connections, or by neurons with predominantly long connections {\it and} neurons with predominantly short connections, and whether they are organized in a columnar structure or not.
From this perspective, the columnar structures may express a sufficient (but not necessary) facilitating structure of a combined speed of information / minimal wiring length optimization. In fact, whereas most monkeys, carnivores and ungulates do have columns, rats and mice don't have them (discounting the dedicated structure of the barrel cortex). The characteristics of the optimal networks (mean path lengths and clustering coefficients) emerging in our study are consistently close to those of the biological examples (e.g., \textit{C. elegans} \cite{Strogatz}). 
%, ruling out that our findings could be the consequence of particular modeling choices. 
Moreover, our excitatory-inhibitory power law decay exponents of the network connectivity ($2$ and $1.5$, respectively) are close to those obtained from 
a critical avalanche model of cortical computation \cite{Arcangelis} for avalanche time duration and avalanche size, respectively. The question emerges whether there is a direct link between our approach and models of cortical computation at criticality \cite{Arcangelis2,Eguiluz}. Detailed numerical experiments along our framework may reveal the nature of this correspondence.

 Since there is no indication of a phase-transition at a finite system size, we expect that the building of large-scale neural networks based on simple electronic neurons arranged according to the physiological template %(a paradigm of the 'blue brain project \cite{bluebrain}) 
will corroborate the two main cortical network features extracted above: An increased speed of information transfer and synchronizability at minimal wiring length.

\vspace{-0.5cm}
\small{

}
\end{document}